\documentclass[aps,prl,twocolumn,floats,showpacs,floatfix]{revtex4}
\usepackage{graphicx} 

\begin{document}
\newcommand{\remark}[1]{{(\it#1)}}
\newcommand{\eexp}{\mbox{e}^}
\newcommand{\new}[1]{{\bf #1}}
\newcommand{\bra}[1]{\left\langle  #1\right|}
\newcommand{\inprod}[2]{\left\langle #1 \vert #2 \right\rangle }
\newcommand{\ket}[1]{\left|#1\right\rangle }
\newcommand{\acos}[1]{\mbox{a}\cos }
\newcommand{\abs}[1]{\vert #1 \vert}
\newcommand{\avg}[1]{\left\langle #1 \right\rangle}
\newcommand{\csch}{\mbox{csch}}

\title{Probing Localization in Absorbing Systems via Loschmidt Echos}

\author{Joshua D. Bodyfelt$^{1}$, Mei C. Zheng$^{1,2}$, Tsampikos Kottos$^{1, 2}$, Ulrich Kuhl$^{3}$, and
Hans-J\"{u}rgen St\"{o}ckmann$^{3}$}

\affiliation{$^1$Department of Physics, Wesleyan University, Middletown, Connecticut 06459, USA \\
$^2$MPI for Dynamics and Self-Organization - Bunsenstra\ss e 10, D-37073 G\"{o}ttingen, Germany \\
$^3$Fachbereich Physik, Philipps-Universit\"{a}t Marburg, Renthof 5, D-35032 Marburg, Germany}

\begin{abstract}
We measure Anderson localization in quasi-one-dimensional waveguides in the presence of absorption by
analyzing the echo dynamics due to small perturbations. We specifically show that the inverse participation
number of localized modes dictates the decay of the Loschmidt echo, differing from the Gaussian decay expected
for diffusive or chaotic systems. Our theory, based on a random matrix modeling, agrees perfectly with
scattering echo measurements on a quasi one-dimensional microwave cavity filled with randomly distributed
scatterers.
\end{abstract}

\pacs{42.25.Dd, 72.15.Rn, 03.65.Nk, 03.65.Yz}

\maketitle

The propagation of waves through complex media is an interdisciplinary problem that addresses areas as
diverse as light propagation in fog or clouds, to electronic and atomic-matter waves used to transmit
energy and information. Despite this diversity, the wave character of these systems provides a common
framework for understanding their transport properties and often leads to new applications. One such
characteristic is a wave interference phenomenon. Its existence results in a complete halt of wave
propagation in random media which can be achieved by increasing the randomness of the medium. This
phenomenon was predicted fifty years ago in the framework of quantum (electronic) waves by Anderson
\cite{A58} and since then has developed as a field of its own.

\begin{figure}
\hspace*{-1cm}\mbox{\includegraphics[width=\columnwidth,height=6cm,clip]{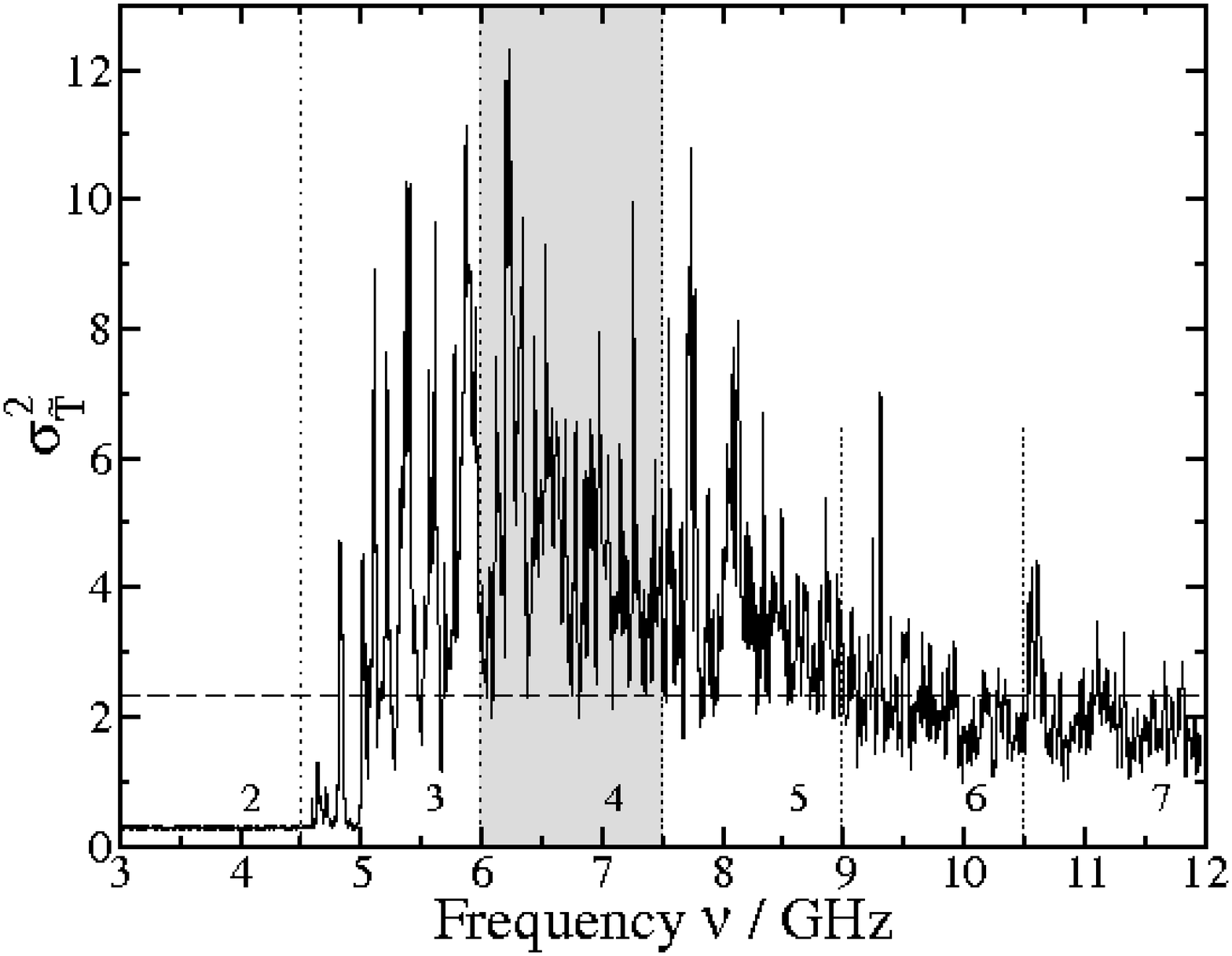}
{\hspace*{-7.85cm}\raisebox{1.8cm}{\includegraphics[height=3.9cm,clip]{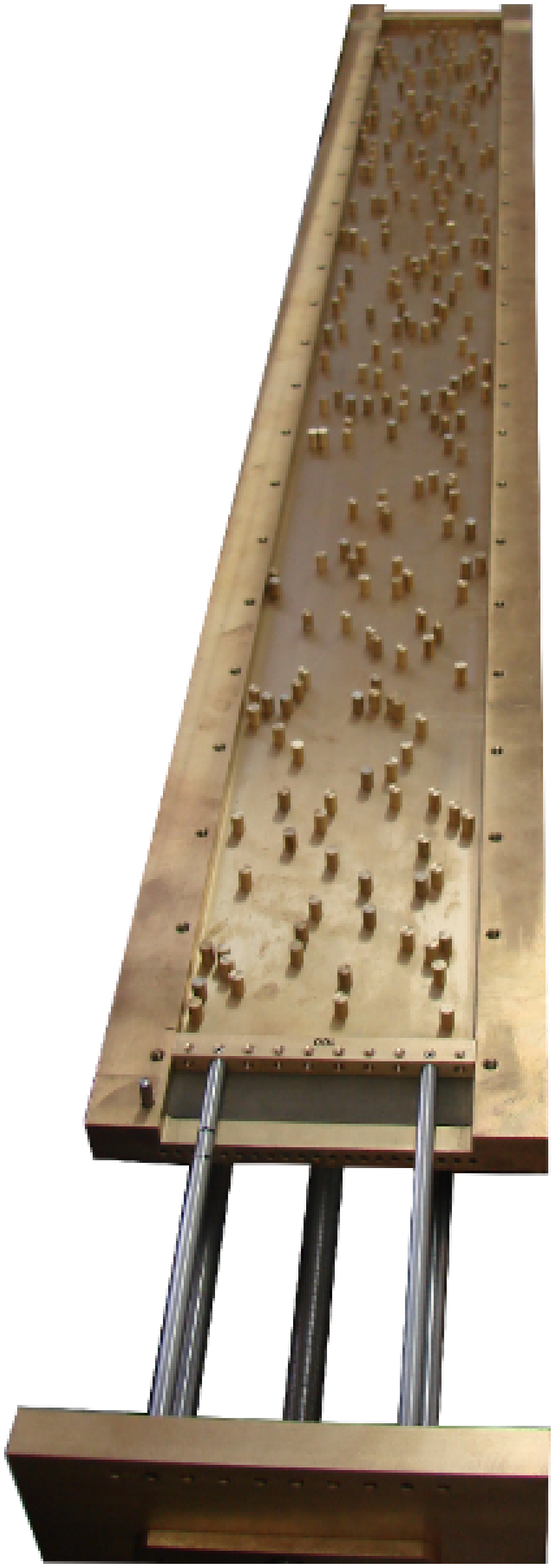}}}\hspace*{2.8cm}\raisebox{3.3cm}{\includegraphics[width=3.2cm,clip]{fig1b}}}
\caption{Left Inset: Experimental apparatus - the top plate with two mounted antennae has been removed to show the waveguide with scatterers. Main Figure: The variance of the normalized transmission intensity $\tilde T = \abs{S_{21}}^2 / \avg{\abs{S_{21}}^2}$. Vertical dashed lines are the mode cut-off frequencies where the number between is the number of open modes. The horizontal dashed line is the critical  localization threshold of $7/3$. The shaded region is a localized frequency window of $6.0-7.5$\,GHz. Right Inset: Normalized transmission distribution ${\cal P}({\tilde T})$ for the localized frequency window of $6.0-7.5$\,GHz. The red dashed line is a fit of the core to a log-normal distribution.}
\label{fig:var}
\end{figure}

Despite the enormous research efforts by various groups in measuring Anderson localization, it took
nearly 40 years to observe localization phenomena beyond any doubt. A decisive step towards this
direction was done by optics and microwave experiments which allow a detailed study of the Anderson
localization, undisturbed by interactions or other effects which characterize electron propagations.
First experiments showing photon localization \cite{rae89} had the problem of separating localization
from absorption, which can be another source of exponential decay of a propagating electromagnetic wave.
A solution to this problem was given by Chabanov et.\,al in Ref.~\cite{CSG00}, where they proposed to
study the relative size of fluctuations of certain transmission quantities. They found clear evidence
of localization in a quasi-one-dimensional (1D) microwave waveguide with randomly distributed dielectric or
metallic spheres \cite{CSG00} (see also Fig.~\ref{fig:var}).

This approach, although quite powerful, does not allow the view of transport from a dynamical perspective,
nor makes a direct contact with the original ideas of Anderson theory, which suggests probing localization
by means of the sensitivity of the system properties against small perturbations \cite{edw72}. This approach
led us in recent years to focus on new measures that efficiently probe the complexity of quantum time evolution.
One such measure is the so-called Loschmidt echo (LE), or fidelity, which probes the sensitivity of quantum
dynamics to external perturbations (for review, see Ref. \cite{GPSZ06}). The recent literature on the subject
is vast and ranges from atomic physics \cite{GCZ97}, microwaves \cite{SGSS05}, elastic
waves \cite{LW03} to quantum information \cite{NC00}, and quantum chaos \cite{JP01,JSB01,CT02,PS02,BC02}.
Formally, the LE $F^{\lambda}(t)$, is defined as \cite{SGSS05}
\begin{equation}
\label{eq:Ft}
F^{\lambda}(t) \equiv |f^{\lambda}(t)|^2 = |\bra{\psi_0}\eexp{iH_0t/\hbar}\eexp{-iH_{\lambda}t/\hbar}\ket{\psi_0}|^2 ;
\end{equation}
where $f^{\lambda}(t)$ is the fidelity amplitude, $H_{\rm \lambda}=H_{\rm 0}+\lambda V$ is a one-parameter
family of Hamiltonians, $H_0$ is the unperturbed Hamiltonian, $\lambda V$, where $\langle\left|V_{nm}\right|^2\rangle=1$,
represents a perturbation of strength $\lambda$, and $\ket{\psi_0}$ is an initial state.

Fidelity in its original definition, Eq.~(\ref{eq:Ft}), is hardly accessible to any experiment where the
information about the system's state is based on the measurements of certain observables; the most
popular being the scattering matrix itself. Therefore, the notion of scattering fidelity had been
introduced as an alternative to Eq.~(\ref{eq:Ft}) \cite{sch05b}. In fact, it was shown that under
certain conditions the scattering fidelity coincides with the standard fidelity \cite{sch05b}.

We present here the first measurements of fidelity of localized waves. Using a quasi-1D
disordered cavity in the localized regime (see left inset of Fig.~\ref{fig:var}), we investigate the fidelity
decay of microwave radiation, due to small perturbations $\lambda$ in the form of boundary displacements of the
sample, and find deviations from a Gaussian decay expected in a frequency interval associated with
extended waves. Using a banded random matrix theory (RMT) modeling, we find that the fidelity amplitude
decays as:
\begin{equation}
\label{eq:LocalTheory}
f(t) \simeq  (\alpha t)^2 \csch^2(\alpha t) ;\quad \alpha=\lambda \sqrt {1.5 I_2}
\end{equation}
where $I_2=\int |\psi({\bf r})|^4 d{\bf r}$ is the inverse participation number (IPN). Using a scaling
analysis of $\alpha$ with respect to $\lambda$ we extract $I_2$ and measure the localization properties
of the sample, {\it even if absorption is present}. Our theoretical results, Eq.~(\ref{eq:LocalTheory}), are
confirmed by our experimental measurements of scattering fidelity.

The scattering fidelity amplitude is defined as
\begin{equation}
\label{fab}
f^\lambda_{ab}(t)=\langle S_{ab}^{\lambda *}(t)S^0_{ab}(t)\rangle\left/
\sqrt{\langle\left|S^\lambda_{ab}(t)\right|^2\rangle \langle\left|S^0_{ab}(t)\right|^2\rangle}\right.
\end{equation}
where $S_{ab}(t)$ is one component of the scattering matrix in the time domain, while the indices $a$
and $b$ refer to the antennae involved. In microwave studies the $S_{ab}$ are directly accessible from
transmission or (for $a=b$) reflection measurements in the frequency domain. The corresponding quantities
in the time domain are then obtained by Fourier transforms. The super-indices $\lambda, 0$ indicate
scattering matrix elements corresponding to the perturbed and unperturbed system, respectively. It is
important to point out that the denominator in Eq.~(\ref{fab}) renormalizes the fidelity decay due to
absorption, thus allowing us to trace out localization phenomena \cite{sch05b}.

The experiment has been performed in a quasi-1D rectangular waveguide (height $8$\,mm, width $100$\,mm, 
length $1190$\,mm) containing $186$ randomly distributed brass cylinders of radius $5$\,mm
(see left inset of Fig.~\ref{fig:var}). One end of the waveguide holds a fixed reflecting metallic wall,
while at the other end there is a moveable reflecting metallic wall, which can be adjusted by means of
a step motor. One antenna was placed close to the moving wall, while another one was placed in the center
of the scattering arrangement. Measurements have been performed in the frequency range $3$ to $12$\,GHz. 
The cut-off frequency for the lowest mode is $1.5$\,GHz, while at the upper limit of the studied frequency range there are $7$
propagating modes. The reflection amplitude at the center antenna ($S_{22}$) and the transmission amplitudes
between the two antennae ($S_{21}$) have been measured for different wall positions. The reflection amplitude
$S_{22}$ is used in Eq.~(\ref{fab}), and the transmission amplitudes $S_{21}$ are used in transmissive studies
below. The wall shift has been performed in steps of $\delta w=0.2$\,mm up to a total shift of $18$\,mm.
In addition, an ensemble average over $15$ different realizations of scatterer positions has been performed.

In order to get confidence that our analysis is performed within the appropriate frequency window where
localization is present, we first investigate the variance $\sigma_{{\tilde T}}^2$ of the normalized
transmission intensity ${\tilde T}=\left|S_{21}\right|^2/\langle \left|S_{21}\right|^2\rangle$
~\cite{CSG00}. Since our experiment does not probe the total transmission but just one component of the
scattering matrix, we expect localization whenever $\sigma_{{\tilde T}}^2$ exceeds the critical value
of $7/3$ \cite{CSG00}. We find (see Fig.~\ref{fig:var}) that this condition is satisfied approximately
in the frequency window $5.5-9$\,GHz. Above $9$\,Ghz the waveguide modes are delocalized, while below
$5.5$\,GHz the values of the variances are error prone, as $S_{21}$ is below the precision of the
vector network analyzer ($|S_{21}| < 10^{-6}$). In the delocalized regime random matrix theory predictions
are applicable \cite{bar94}, yielding a value of $\sigma_{{\tilde T}}^2 =(2N+1)^2/[N(2N+3)]-1$, where $N$ is the
number of open channels. In the limit $N\to\infty$, the variance approaches the value $\sigma_{{\tilde T}}^2=1$,
in agreement with our experimental data for the high frequency regime (see Fig.~\ref{fig:var}).

We have also investigated the whole normalized transmission distribution ${\cal P}({\tilde T})$. As expected
\cite{M00}, we find a transition from a Rayleigh-like behavior (applicable in the delocalized regime) to a
broader distribution approaching a log-normal behavior deep in the localized regime (see right inset of
Fig.~\ref{fig:var} showing ${\cal P}({\tilde T})$ in the localized frequency window of $6-7.5$\,GHz.).

\begin{figure}
\includegraphics[width=\columnwidth,height=6cm,clip]{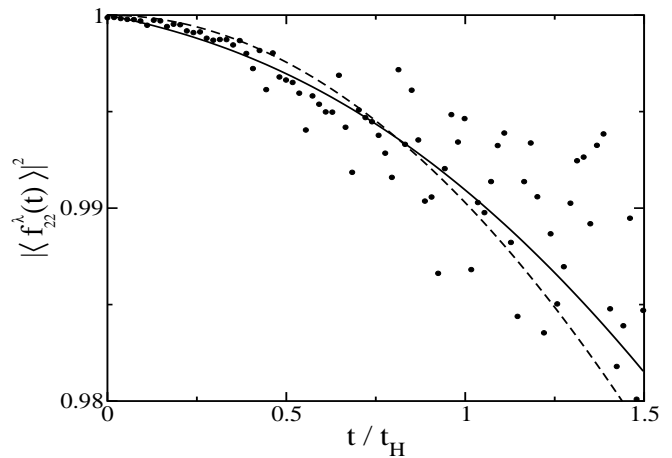}
\caption{Experimental scattering fidelity in the frequency window of 6 to 7.5\,GHz
(localized). The dots correspond to the experimental scattering fidelity, Eq.~(\ref{fab}),
for a wall shift of $0.8$\,mm. The dashed line is a best fit to Eq.~(\ref{eq:GOE}), while the
solid line is a best fit to Eq.~(\ref{eq:LocalTheory}). Both fits were done for the region
$t \leq t_H$. Eq.~(\ref{eq:LocalTheory}) can be seen as the better fit.}
\label{fig:ExpFid}
\end{figure}

Next, we measure the scattering fidelity in the localized frequency window (see previous analysis), and
compare the experimental data with the LE decay law found for chaotic/diffusive systems \cite{SGSS05,
GPSZ06}. The latter reads
\begin{equation}
\label{eq:GOE}
F(t) \simeq e^{ -\lambda^2 {\cal C}(t)}; \,\,
{\cal C}(t)\equiv t^2 + t - \int_0^t d\tau \int_0^{\tau} d\tau' b_2(\tau'),
\end{equation}
which for small perturbations can be approximated as $F(t)\sim \exp(-(\lambda t)^2)$. Above, $b_2$ is
the two-point form factor for the Gaussian orthogonal ensemble \cite{GPSZ06}. Using
$\lambda$ as a fitting parameter, we have attempted to fit the experimental data with Eq.~(\ref{eq:GOE}).
We found that the overall agreement is poor (see Fig.~\ref{fig:ExpFid}). Further analysis (see below)
confirms that Eq.~(\ref{eq:GOE}) is inapplicable in the localized regime. On the contrary, when fitting
the experimental data with Eq.~(\ref{eq:LocalTheory}) we get an excellent agreement.

The analytical calculation of $f(t)$ relies on the relation between scattering fidelity and LE in the
weak coupling regime where our experiment operates. The LE was further evaluated using a RMT modeling
for $H_0$ and $V$. For diffusive or chaotic cavities, $H_0$ and $V$ are modeled by matrices
drawn from a Gaussian Orthogonal Ensemble (GOE). Anderson localization in quasi-1D
disordered systems, on the other hand, is modeled by sharp banded GOE matrices with bandwidth $b$ for $H_0$ and
$V$ \cite{IKPT97,FM94}. In this model, the localization length scales as $l_{\infty}\sim b^2$.
Therefore localization sets in for matrices of rank $L\gg b^2$.

Expanding the initial preparation as $|\psi_0\rangle=\sum_n a_n|\phi_n^0\rangle$, where $H_0|\phi_n^0
\rangle = E_n^0 |\phi_n^0\rangle$, it is found that the fidelity amplitude defined in Eq.~(\ref{eq:Ft})
can be written as
\begin{equation}
\label{eq:fid}
f(t) = \sum_{n,m,k} \alpha_n^\ast \alpha_k \inprod{\phi_m^{\lambda}}{\phi_k^0} \inprod{\phi_n^0}
{\phi_m^{\lambda}} e^{-i (E_m^{\lambda} - E_n^0) t}
\end{equation}
where $H_{\lambda}|\phi_n^{\lambda}\rangle = E_n^{\lambda}|\phi_n^{\lambda}\rangle $. Averaging over
disordered realizations one further gets that
\begin{equation}
\label{finalf1}
\overline{f(t)} =  \sigma \overline{ \sum_{n,m} {\cal L}_{mn} \exp[-i (E_m^{\lambda} - E_n^0) t]}
\end{equation}
where $\sigma = {1\over l_\infty}\sum_{ n \leq l_{\infty} }\overline{\abs{\psi_0(n)}^2}$, $\psi_0(n)$
are the components of the initial preparation $|\psi_0\rangle$ in the position (Wannier) basis, and ${\cal
L}_{mn} = \abs{\inprod{\phi_m^{\lambda}}{\phi_n^0}}^2$ is the Local Density of States (LDoS) kernel
\cite{CK01}.
In order to derive Eq.~(\ref{finalf1}) we have assumed statistical independence between the eigenfunctions
and eigenvalues of our Hamiltonian, while localization enforces the following contraction rule
\begin{equation}
\overline{\alpha_n^\ast \alpha_k} = \sigma \; \delta_{n,k}.
\end{equation}
\begin{figure}
\includegraphics[width=\columnwidth,height=6cm,clip]{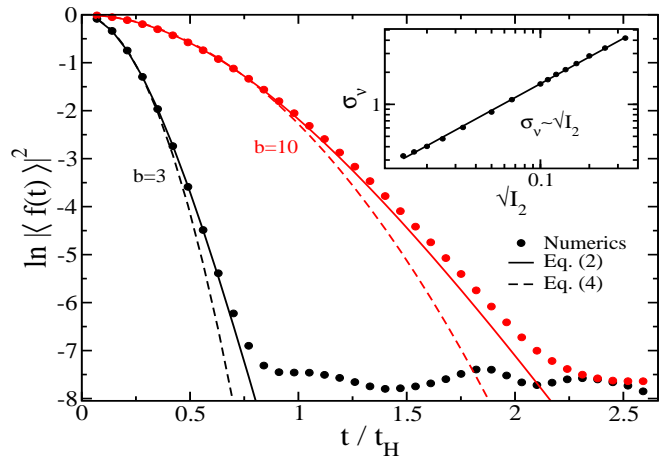}
\caption{Numerical fidelity, Eq.~(\ref{eq:Ft}), for Hamiltonians modeled by banded random matrices.
The parameters are $\lambda = 10^{-3}$, $L = 5000$, $b = 10$ (upper curve), and $b=3$ (lower curve).
Circles are the numerical results. Dashed lines are best fits to Eq.~(\ref{eq:GOE}). Solid lines
are best fits to Eq.~(\ref{eq:LocalTheory}), which again show a better fit. The inset shows the variance extracted
from the fit of Eq.~(\ref{eq:LocalTheory}), plotted against the root of the inverse participation number. The
observed linear relation confirms the validity of the theoretical calculations.}
\label{fig:NumFidLoc}
\end{figure}

For small enough perturbations, such that only nearby (on the order of mean level spacing) energy levels
are mixed, one can approximate ${\cal L}_{mn}$ in Eq.~(\ref{finalf1}) with a delta-function. The fidelity
amplitude is then written as
\begin{equation}
\label{finfidf}
\overline {f(t)} \simeq \overline{\sum_n\exp(-i v_n \lambda t)}=\int dv {\cal P}(v) \exp( - i v \lambda t)
\end{equation}
where $v_n=(E_n^{\lambda}-E_n^0)/\lambda$ are the so-called level velocities, and ${\cal P}(v)$ is their
corresponding distribution. The latter has been calculated in Ref.~\cite{F94}, and it was found that in
the localized regime it is given by the expression
\begin{equation}
\label{eq:LLV}
P(\eta) = \frac{\pi}{6} \frac{\pi \eta \coth(\pi \eta / \sqrt{6} ) - \sqrt{6}}{\sinh^2(\pi \eta / \sqrt{6})},
\end{equation}
where $\eta = v / \sigma_{v}$. For localized level velocities, the variance is $\sigma_{v} = \sqrt{I_2}$
\cite{FM95}. Substituting the above distribution in Eq.~(\ref{finfidf}) we finally get the expression given
in Eq.~(\ref{eq:LocalTheory}). The latter is checked numerically for two different bandwidths $b=10, 3$ and for a
system size $L=5000$. Our numerical results are reported in Fig.~\ref{fig:NumFidLoc}, together with
Eqs.~(\ref{eq:LocalTheory}, \ref{eq:GOE}). To confirm further the validity of our calculations, we have fit
the decay of LE for various bandwidths $b$, with Eq.~(\ref{eq:LocalTheory}). From the fit we have extracted
$\sigma_{v}$, which we have plotted against the IPN $I_2$, in the inset of Fig.~\ref{fig:NumFidLoc}. The
observed linear behavior gives further confirmation to our theoretical calculations.

We then analyze the decay of scattering fidelity due to small displacements of one wall of the
disordered quasi-1D waveguide shown in the inset of Fig.~\ref{fig:var}. The shift of the wall
can be mapped onto an effective Hamiltonian $H_{\lambda_w}$ with matrix elements \cite{sch05b}
\begin{equation}\label{heff}
\left(H_{\lambda_w}\right)_{nm}=-w\int\limits_0^L\nabla_\perp[\psi_n(y)]\nabla_\perp[\psi_m(y)]\,dy
\end{equation}
where $w$ and $L$ are the shift and length of the moving wall, respectively, and $\nabla_\perp\psi_n$ and
$\nabla_\perp\psi_m$ are the normal derivatives of the wave functions at the wall. Thus
$\sigma_{\lambda_w}\propto w^2$. The proportionality constant is $(2Lk^3/3\pi^3)$, for the case of 
chaotic cavities in the semiclassical limit \cite{sch05b}. In any case, we finally get that $\lambda_w\sim w$. 

\begin{figure}
\includegraphics[width=\columnwidth,height=6cm,clip]{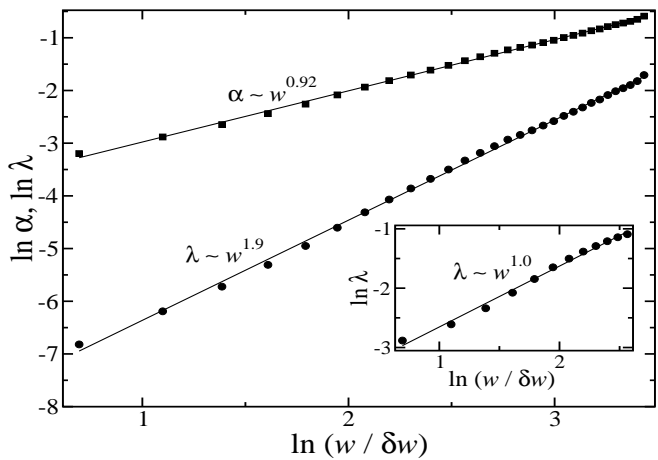}
\caption{Within the localized frequency window of $6-7.5$\,GHz, the experimental fitting parameters
 - $\alpha$ (squares) from Eq.~(\ref{eq:LocalTheory}) and $\lambda$ (circles) from Eq.~(\ref{eq:GOE}) 
- are plotted against the rescaled wall shift $w/\delta w$. Straight lines indicate best fit to power laws, 
$\alpha\sim w^{0.92}$, $\lambda\sim w^{1.9}$.
Inset: Same as main figure, but for the delocalized frequency window of $10.5-12$\,GHz. 
As opposed to the main figure, here $\lambda\sim w^{1.0}$, in agreement with Eq.~(\ref{eq:GOE}).}
\label{fig:ExpFidLoc}
\end{figure}


Next we proceed by fitting our experimental data on scattering fidelity with Eqs.~(\ref{eq:LocalTheory}, 
\ref{eq:GOE}) where $\alpha, \lambda$ are used as fitting parameters. We have extracted $\alpha, \lambda$, 
for various wall shifts $w$ and plotted them versus a rescaled shift, $w/\delta w$. The results are summarized
in Fig.~\ref{fig:ExpFidLoc}. We find that in the frequency window $6-7.5$\,GHz (where Anderson localization
is present), the best fit with Eq.~(\ref{eq:GOE}) gives $\lambda\sim \lambda_w^{\gamma}$, with $\gamma
=1.9 \pm 0.05$. This result violates the theoretical expectation $\gamma\approx 1$ and constitutes a
direct confirmation that the RMT result Eq.~(\ref{eq:GOE}) is not applicable in the localized regime.
At the same time, the best fit of the experimental scattering fidelity with the prediction of the banded
RMT modeling (\ref{eq:Ft}) gives that $\alpha\sim \lambda_w^{\gamma}$ with $\gamma\sim 0.92 \pm 0.05$,
in agreement with our theory. The extracted slope $\alpha/\lambda_w \sim {\sqrt I_2}$ can be used as an 
estimation for the localization properties of our sample. Within the delocalized frequency window $10.5
-12$\,GHz, a fit with Eq.~(\ref{eq:GOE}) works perfectly well with $\gamma\approx 1.0\pm 0.05$, in nice 
agreement with theory (see inset of Fig.~\ref{fig:ExpFidLoc}).

In conclusion, using echo dynamics, we were able to isolate absorption phenomena and identify unambiguously
traces of localization in random media. Our theoretical calculations, based on a RMT modeling, indicated
that the IPN of localized modes dictates the behavior of LE which follows a novel decay law; being
experimentally distinguishable from the Gaussian decay observed in diffusive/chaotic systems. Our
experimental measurements with disordered waveguides confirm the theoretical expectations, thus
suggesting fidelity as a reliable tool to investigate localization in the presence of absorption.

The research was funded by the Deutsche Forschergruppe 760 ``Scattering Systems with Complex Dynamics'' and by
a grant from the United States-Israel Binational Science Foundation (BSF), Jerusalem, Israel.



\begin{thebibliography}{99}

\bibitem{A58} P.~Anderson, Phys. Rev. {\bf 109}, 1492 (1958).

\bibitem{rae89} H.~D. Raedt, A.~Lagendijk, P.~Vries, Phys. Rev. Lett. {\bf 62}, 47 (1989).

\bibitem{CSG00} A.~Chabanov, M.~Stoytchev, A.~Genack, Nature {\bf 404}, 850 (2000); A.~Chabanov,
A.~Genack, Phys. Rev. Lett. {\bf 87}, 233903 (2001); A.~Genack, A.~Chabanov, J. Phys. A \textbf{38},
10465 (2005).

\bibitem{edw72} J.~T. Edwards, D.~J. Thouless,
  \bibinfo{journal}{J. Phys. Chem.} \textbf{\bibinfo{volume}{5}},
  \bibinfo{pages}{807} (\bibinfo{year}{1972}).

\bibitem{GPSZ06} T. Gorin, et al., Phys. Rep. {\bf 435}, 33 (2006);
Ph. Jacquod, C. Petitjean, arXiv:0806.0987v1 (2008).

\bibitem{GCZ97} S.~A. Gardiner, J. I. Cirac and P. Zoller, Phys. Rev. Lett. {\bf 79}, 4790 (1997); 
M.~F. Andersen, A. Kaplan and N. Davidson, Phys. Rev. A {\bf 64}, 043801 (2001); S. Kuhr et al. Phys. 
Rev. Lett. {\bf 91}, 213002 (2003).

\bibitem{SGSS05} R.~Sch\"afer, et al.,
  \bibinfo{journal}{Phys. Rev. Lett.} \textbf{\bibinfo{volume}{95}},
  \bibinfo{pages}{184102} (\bibinfo{year}{2005}).

\bibitem{LW03} O. Lobkis, R. Weaver, Phys. Rev. Lett. {\bf 90}, 254302 (2003).

\bibitem{NC00} M.~A. Nielsen, I.~L. Chuang, \emph{Quantum computation and quantum information}
(Cambridge University Press,2000).

\bibitem{JP01} R.~A. Jalabert and H.~M. Pastawski, Phys. Rev. Lett. {\bf 86}, 2490 (2001);
F.~M. Cucchietti, H.~M. Pastawski and D.~A. Wisniacki, Phys. Rev. E {\bf 65}, 046209 (2002).

\bibitem{JSB01} Ph. Jacquod, I. Adagdeli and C.~W.~J. Beenakker, Phys. Rev. Lett. {\bf 89},
154103 (2002); Ph. Jacquod, I. Adagideli and C.~W.~J. Beenakker, Europhys. Lett.{\bf 61}, 729 (2003).

\bibitem{CT02} N.~R. Cerruti and S. Tomsovic, Phys. Rev. Lett. {\bf 88}, 054103 (2002);
J. Phys. A: Math. Gen. {\bf 36}, 3451 (2003).

\bibitem{PS02} T. Gorin, T. Prosen, T. H. Seligman, New J. Phys. {\bf 6}, 1 (2004); T. Prosen, 
M. Znidaric, J. Phys. A: Math. Gen. {\bf 35}, 1455 (2002); T. Kottos, D. Cohen, Europhys. Lett. 
{\bf 61}, 431 (2003); M. Hiller, et al., Phys. Rev. Lett. {\bf 92}, 010402 (2004).

\bibitem{BC02} G. Benenti and G. Casati, Phys. Rev. E {\bf 66}, 066205 (2002); Y. Adamov, I.~V.
Gornyi and A.~D. Mirlin, Phys. Rev. E {\bf 67}, 056217 (2003); G. S. Ng, J. Bodyfelt and T. Kottos,
Phys. Rev.  Lett {\bf 97}, 256404 (2006).

\bibitem{sch05b} R.~Sch\"{a}fer, et al.,
\bibinfo{journal}{New J. of Physics}
  \textbf{\bibinfo{volume}{7}}, \bibinfo{pages}{152} (\bibinfo{year}{2005}).

\bibitem{bar94} H.~U. Baranger, P.~A. Mello, \bibinfo{journal}{Phys. Rev. Lett.} \textbf{\bibinfo{volume}{73}},
  \bibinfo{pages}{142} (\bibinfo{year}{1994}).

\bibitem{M00} A. Mirlin, Phys. Rev. E {\bf 57}, R6285 (1998); A. A. Chabanov, A. Z. Genack, ibid.
{\bf 72}, 055602(R) (2005); M. Stoytchev, A. Z. Genack, Phys. Rev. Lett. {\bf 79}, 309 (1997).

\bibitem{IKPT97} G.~Casati, et al.,
  \bibinfo{journal}{Phys. Rev. Lett.} \textbf{\bibinfo{volume}{72}},
\bibinfo{pages}{2697} (\bibinfo{year}{1994});
F.~Izrailev, et al.,
\bibinfo{journal}{Phys. Rev. E} \textbf{\bibinfo{volume}{55}},
  \bibinfo{pages}{4951} (\bibinfo{year}{1997}).

\bibitem{FM94} Y.V.Fyodorov, A.D.Mirlin,
\bibinfo{journal} {Int. J. Mod. Phys.} \textbf{\bibinfo{volume}{8}},
   \bibinfo{pages}{3795} (\bibinfo{year}{1994}).

\bibitem{CK01} D.~Cohen, T.~Kottos,
  \bibinfo{journal}{Phys. Rev. E} \textbf{\bibinfo{volume}{63}},
  \bibinfo{pages}{036203} (\bibinfo{year}{2001}); M. Hiller, T. Kottos, T. Geisel, Phys. Rev. A {\bf 73}, 061604 (2006).

\bibitem{F94} Y.~V.~Fyodorov,
  \bibinfo{journal}{Phys. Rev. Lett.} \textbf{\bibinfo{volume}{73}},
  \bibinfo{pages}{2688} (\bibinfo{year}{1994}); P.~Kunstman, K.~Zyczkowski, J.~Zakrzewski,
  \bibinfo{journal}{Phys. Rev. E} \textbf{\bibinfo{volume}{55}},
  \bibinfo{pages}{2446} (\bibinfo{year}{1997}).

\bibitem{FM95} Y.~V.~Fyodorov, A.~Mirlin,
  \bibinfo{journal}{Phys. Rev. B} \textbf{\bibinfo{volume}{51}},
  \bibinfo{pages}{13403} (\bibinfo{year}{1995}).
\end{thebibliography}
\end{document}